\documentclass[10pt,letterpaper]{article}
\usepackage[cmex10]{amsmath}
\usepackage{opex3}
\usepackage{array}
\usepackage{mdwmath}
\usepackage{mdwtab}
\usepackage{subeqnarray}
\usepackage{cases}
\usepackage{cite}
\usepackage{float}
\usepackage{graphicx}
\usepackage{caption}
\usepackage{subcaption}

\begin{document}

\title{Cylindrical Beam Propagation Modelling of Perturbed Whispering-Gallery Mode Microcavities}

\author{Mohammad~Amin~Cheraghi~Shirazi, Wenyan~Yu, Serge~Vincent,~and~Tao~Lu$^*$}

\address{Department of Electrical and Computer Engineering, University of Victoria, EOW 448, 3800 Finnerty Rd., Victoria, BC, V8P 5C2, Canada.}
\email{$^*$taolu@ece.uvic.ca}
\homepage{http://www.ece.uvic.ca/{\textasciitilde}taolu} 

\begin{abstract}
We simulate light propagation in perturbed whispering-gallery mode microcavities using a two-dimensional finite-difference beam propagation method in a cylindrical coordinate system. Optical properties of whispering-gallery microcavities perturbed by polystyrene nanobeads are investigated through this formulation. The light perturbation as well as quality factor degradation arising from cavity ellipticity are also studied.
\end{abstract}

\ocis{(230.3990) Micro-optical devices; (040.1880) Detection; (230.5750) Resonators.}


\section{INTRODUCTION}
Whispering-gallery mode (WGM) microcavities are at the frontier of research on subjects ranging from biosensing, nonlinear optics, and laser physics, to fundamental physics such as cavity quantum electrodynamics\cite{toroid_Gorodetsky_Ilchenko_Ultimate_Microsphere,toroid_Armani_Ultra,Vollmer_Arnold_Protein_Microcavity,biosensing_Vollmer_Arnold_Whispering_molecules,Lu_Vahala_High_Sensitivity,Dominguez_Martorell_Whispering_molecules,Knittel_Bowen_Interferometric_biosensors,Sun_Fan_Optical_Sensing,Spillane_Vahala_Ultralow_Microcavity,Min_Vahala_Compact_Laser,Cai_Vahala_Highly_Laser,toroid_Polman_Ultralow,Lu_Vahala_On_Chip_Microlaser,Lu_Min_Narrow_Laser,toroid_Spillane_Ultrahigh}. Contrary to its rapid experimental advances, numerical exploration of WGM's has been largely lagging behind with a limited number of available options\cite{toroid_Oxborrow_Traceable,Poon_Yariv_Matrix_Waveguides, TMM,Du:13}.  On the other hand, the Beam Propagation Method (BPM) has a long history\cite{Feit_Fleck_Light_Fibers, yevick_1990, yevick1983, huang-march92, vanroey1981,huang1992, yevick1984, Osgood, Gopinath} in modelling light propagation along both straight and curved waveguides as well as whispering-gallery microcavity eigenmode analyses\cite{FE-BPM_bangiladish}. Compared to  boundary element\cite{Yang_Boundary_new, pmd_Lu_Boundary, pmd_Lu_Comparative, pmd_Lu_Vectorial}, finite element\cite{pmd_Deng_Nonunitarity_new}, finite-difference time-domain\cite{FDTD_waterloo}, and free space radiation mode methods\cite{mode1996}, BPM remains highly efficient without sacrificing substantial accuracy.  By adopting a perfectly matched layer (PML)\cite{PML_Berenger1994, PML_Huang} to absorb the light which is otherwise reflected at the computation window and following the procedure formulated in \cite{hadley2002} to correct inaccuracies incurred at the refractive index discontinuities in high refractive index contrast waveguide structures, the Finite-Difference Beam Propagation Method (FD-BPM) can achieve high accuracy with a rapid convergence rate. Conventional FD-BPM formulations are based on the Fresnel approximation, where light is assumed to propagate close to the propagation axis\cite{vanroey1981, yevick1983,yevick1984 , huang-march92}. To overcome this limitation for bent waveguide modelling, high-order algorithms known as wide-angle BPM\cite{Wide_angle_Hadley-Pade} or the conformal mapping approach\cite{cyl-FD-BPM} are desirable. Alternatively, BPM may be reformulated in cylindrical coordinates systems to analyze such structures\cite{rivera1995, 3D-curved, krause2011}.

\begin{figure}[b]
  \centering
  \includegraphics[width=0.45\textwidth]{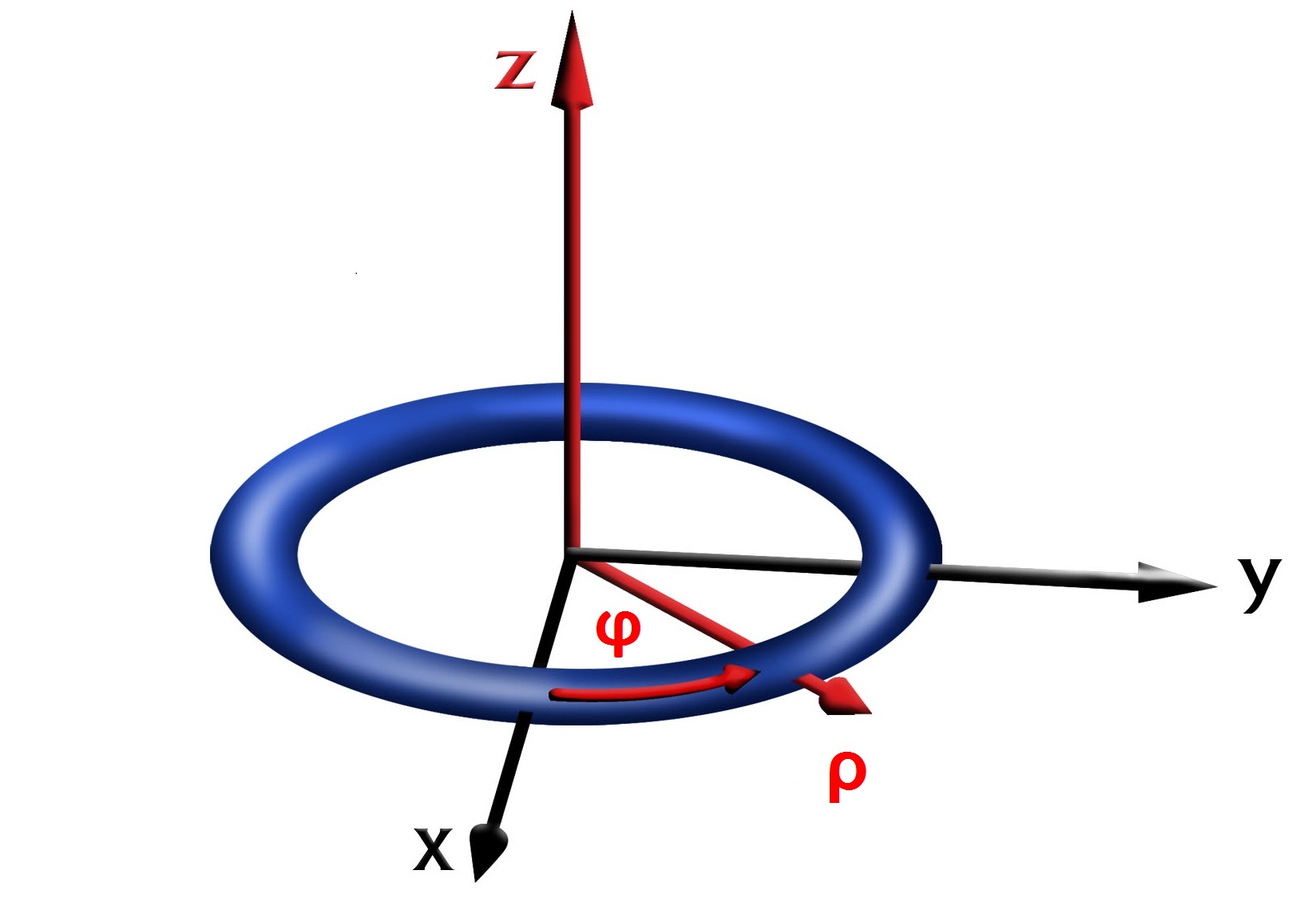}
  \caption{A cylindrical coordinate system.}\label{cyl_coor}
\end{figure}

In this work, we simulated the light propagation in a WGM microcavity by implementing FD-BPM in a cylindrical system as shown in Fig.~\ref{cyl_coor}. The field perturbation from a nanobead attached to the microcavity and quality factor degradation arising from cavity deformations were investigated. The computed field distribution correctly includes the radiative field component, which a mode analysis technique would fail to simulate.

\section{\label{sec:level3}Formulation}

From the Helmholtz equation, the field component $E_z$ of the TE wave in a two-dimensional whispering-gallery microcavity satisfies

\begin{equation}\label{wave_equation}
\frac{\partial^2 E_z}{\partial \phi^2} + \rho^2 \frac{\partial^2 E_z}{\partial \rho^2} + \rho \frac{\partial E_z}{\partial \rho} + n^2(\rho,\phi)k_0^2 \rho^2 E_z =0
\end{equation}
where $n(\rho,\phi)$ is the refractive index of the cavity, $k_0=2\pi/\lambda$ is the wave number in free space, and $\lambda$ is the vacuum wavelength of the light circulating in the cavity. For a perfect WGM cavity with azimuthal symmetry, the refractive index is independent of $\phi$ ($n(\rho,\phi)=n(\rho)$). The electric field can be approximated as propagating in the form of
\begin{equation}\label{BPM_mode}
E_z(\rho,\phi)=A\hat{\psi}_{m_r}(\rho)\exp{(jm\phi)}
\end{equation}
where $\hat{\psi}_{m_r}(\rho)$ is the normalized $m_r^\text{th}$ azimuthal modal field distribution such that the squared norm $|A|^2$ represents the circulating power of the mode and $m=m_r+jm_i$ is a complex constant whose real part $m_r$ represents the azimuthal mode order when the cavity is in resonance with the circulating light and its imaginary part $m_i$ characterizes the attenuation of the field in the azimuthal direction. Note that, in general, the real part $m_r$ can be any real number for a given wavelength $\lambda$. When a certain wavelength $\lambda_m$ yields an integer value for $m_r$, resonance occurs. In addition, multiple wavelengths may yield an identical integer $m_r$ where eigensolutions ${\hat \psi}_{m_r}$ correspond to resonance whispering-gallery modes with the same azimuthal order $m_r$ but different transverse modes. Both quantities can be obtained from the nonzero solution of the mode equation, in turn described as an eigenequation with eigenvalue $m^2$ derived from Eq.~\eqref{wave_equation}:
\begin{equation}\label{mode_equation}
[\rho^2 \frac{\partial^2 }{\partial \rho^2} + \rho \frac{\partial }{\partial \rho} + n^2(\rho)k_0^2 \rho^2] {\hat \psi}_m =m^2{\hat \psi}_m
\end{equation}
If the aforementioned symmetry is broken due to an azimuthal angle dependent perturbation of the refractive index $n(\rho,\phi)=n(\rho)+\delta n(\rho,\phi)$ where the perturbation $\delta n(\rho,\phi)\ll n(\rho)$, one may reformulate $E_z$ as
\begin{equation}
  E_z(\rho,\phi)=\psi(\rho,\phi)\exp{(j{\bar m}\phi)}
\end{equation}
where ${\bar m}$ is a reference value such that $\psi(\rho,\phi)$ varies slowly along the azimuthal direction or, equivalently, the slowly varying envelope approximation (SVEA) holds. This is mathematically written as
\begin{equation}\label{SVEA}
|\frac{\partial^2 \psi}{\partial \phi^2}| \ll  2  |{\bar m}\frac{\partial \psi}{\partial \phi}|
\end{equation}
It is necessary to point out that the choice of ${\bar m}$ is arbitrary as long as SVEA holds; however, if the wavelength of the light is close to the resonance wavelength of the $m_r^{\text{th}}$ order unperturbed WGM, it is convenient to select ${\bar m}=m$. We will therefore drop the bar in the rest of the text for convenience. Alternatively, one may treat ${\bar m}=m(\phi)$ as a $\phi$-dependent quantity where $m(\phi)$ is obtained from solving Eq.~\eqref{mode_equation} at each angle $\phi$ for higher accuracy. From Eq.~\eqref{SVEA}, we obtain the wave evolution along the azimuthal direction according to
\begin{equation}\label{k_operator}
         -j\frac{\partial}{\partial \phi} \psi =
         \frac{\rho^2}{2m}\frac{\partial^2 \psi}{\partial \rho^2}+ \frac{\rho}{2m} \frac{\partial \psi}{\partial \rho}+ (\frac{\rho^2k_0^2 n^2(\rho,\phi)}{2m}- \frac{m}{2}) \psi
\end{equation}

Discretizing the computation window uniformly so that the coordinates $(\rho_p,\phi_l)$ of each grid $(p,l)$ can be expressed as $\rho_p=p\Delta \rho$, $\phi_l = l \Delta \phi$, and $\psi(\rho_p,\phi_l)=\psi_{p,l}$, one can evolve the field at $\phi_l$ from a previous azimuthal angle $\phi_{l-1}$ according to

\begin{equation}\label{full_eq_3}
\begin{array}{lll}
a_p\psi_{p-1,l+1} + (\frac{1}{j\Delta \phi}+b_p)\psi_{p,l+1} - c_p\psi_{p+1,l+1} =-a_p\psi_{p-1,l}
+ (\frac{1}{j\Delta \phi}-b_p)\psi_{p,l}+ c_p\psi_{p+1,l}
\end{array}
\end{equation}
where
\begin{equation}
\begin{array}{rcl}
a_p&=&\frac{p(1-2p)}{8m}\\
b_p&=&\frac{p^2(2-\Delta \rho^2 k_0^2 n_{p,l+1}^2)}{4m}+\frac{m}{4}\\
c_p&=&\frac{p(1+2p)}{8m}
\end{array}
\end{equation}

Here $\Delta \rho$ and $\Delta \phi$ are grid spacings along the ${\hat \rho}$ and ${\hat \phi}$ directions, as illustrated in Fig.~\ref{cyl_coor}. Also, $n_{p,l}$ is the refractive index of the waveguide structure at each point. Collecting $\psi_{p,l}$ into a ket form ${\mid \psi_{l}\rangle}=(\psi_{p_0,l},\psi_{p_0+1,l},\ldots\psi_{p_0+N,l})^T$ and rearranging Eq.~\eqref{full_eq_3} into a matrix form, we obtain

\begin{equation}\label{matrix_full_eq}
\tilde H  {\mid \psi_{l+1}\rangle} = \tilde D {\mid \psi_{l}\rangle}
\end{equation}
where $\tilde H$ and $\tilde D$ are two tridiagonal matrices. By adopting standard FD-BPM procedures, one may obtain the field evolution via Eq.~\eqref{matrix_full_eq} from the excitation field at $l=0$.

\section{\label{sec:level4}Results and Discussions}

To characterize the BPM, we first tested it on a perfect silica microring resonator immersed in water.  The refractive index of the silica ring was $1.4508 + j(7.11 \times 10^{-12})$ \cite{silica_refract,refractiv_imaginary} at a wavelength of $970~\text{nm}$ and the surrounding water had a refractive index of $1.327 + j(3.37 \times 10^{-6})$ \cite{Water_refract}.

\begin{figure}
   \centering
   \includegraphics[width=\textwidth]{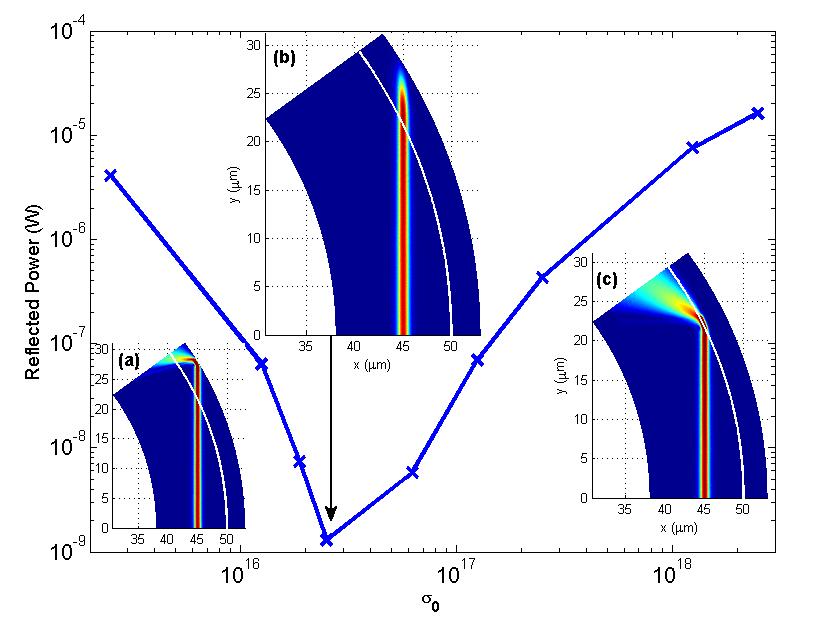}\\
   \caption{Reflected power from a $2$-$\mu m$ PML vs. different $\sigma_0$ values. The insets are the field intensity for three different $\sigma_0$ values: (a) $0$, (b) $2.5 \times 10^{16}$, and (c) $10^{20}$. The white lines indicate the inner PML edges.}\label{PML_Opt}
\end{figure}

 The resonator had a $45$-$\mu m$ major radius and a  $10$-$\mu m$ minor diameter. To simplify the analysis, we reduced the three-dimensional waveguide structure to a two-dimensional one through the use of an effective index method (EIM)\cite{chiang_edd_indx} along the $z$-direction. The cavity was excited by a modal field obtained from the mode solver. To minimize the spurious reflection at the edges of the computation window, a $4$-$\mu m$ PML\cite{PML_Berenger1994} was placed at the edge of the computation window. The PML is implemented by replacing the radial derivative with
\begin{equation}\label{PML}
\frac{\partial}{\partial \rho} \to \frac{1}{1+\frac{j \sigma(\rho)}{\omega}}\frac{\partial}{\partial \rho}
\end{equation}

in which $\sigma(\rho)$ is defined as
\begin{equation}
\sigma(\rho) = \left\{
  \begin{array}{l l}
    \sigma_0(\rho - \rho_0)^2 & \quad \text{Inside the PML}\\
    0 & \quad \text{Elsewhere}
  \end{array} \right.
\end{equation}
where $\rho_0$ is the inner edge of the PML layer and $\sigma_0$ is a constant. To optimize the PML performance and determine the optimal value for $\sigma_0$, a simple experimental simulation was conducted. An optical beam was launched towards a $2$-$\mu m$ PML through a straight waveguide and the reflected power was measured for different values of $\sigma_0$. The different values of reflected power versus $\sigma_0$ are presented in Fig.~\ref{PML_Opt}. The insets show the field intensity for three different $\sigma_0$ values, wherein the white lines are the inner PML edges. The $\sigma_0$ value is 0 in Fig.~\ref{PML_Opt}(a), $2.5 \times 10^{16}$ (i.e. the optimal value) in Fig.~\ref{PML_Opt}(b), and $10^{20}$ in Fig.~\ref{PML_Opt}(c). The optimal value was then utilized for the remaining simulations.

\begin{figure}
  \centering
  \includegraphics[width=0.65\textwidth]{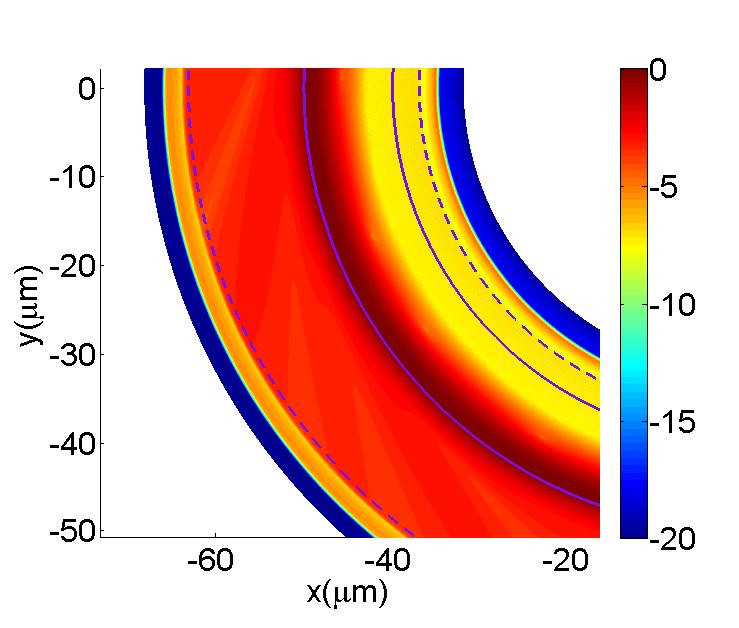}
  \caption{Field intensity in logarithmic scale, where the radiation is observable. To reiterate, the solid blue lines are showing the resonator edges and the dashed blue lines are showing the PML edges.}\label{field_plots}
\end{figure}

In Fig.~\ref{field_plots}, we plot the intensity distribution in logarithmic scale by setting a $36$-$\mu m$ window size, $1601$ grids in the ${\hat \rho}~\text{ direction}$, and $3000$ grids in the ${\hat \phi}\text{ direction}$ for $2\pi~\text{radians}$, where we excited the ring with its fundamental mode. The field distribution is plotted in Fig.~\ref{field_plots}. The resonance wavelength and the mode number calculated by the mode solver are $970.25$~nm and $458$~nm, respectively. The solid blue lines in Fig.~\ref{field_plots} are showing the resonator edges and the dashed blue lines define the PML boundary. As can be seen in this figure, a small portion of the energy radiates towards the computation window edge yet the adopted PML efficiently prevents the otherwise spurious reflection from returning to the resonator.

\begin{figure}
        \centering
        \begin{subfigure}[b]{.49\textwidth}
                \centering
                \includegraphics[width=\textwidth]{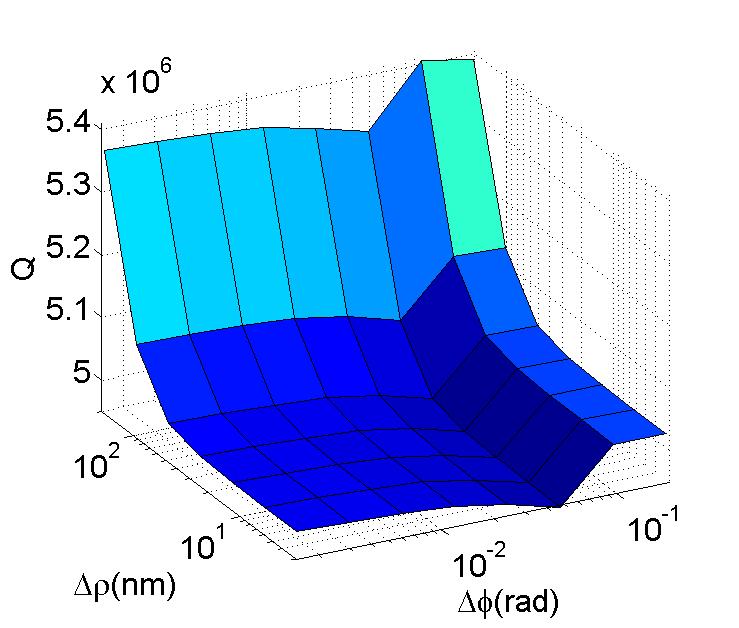}
                \caption{}
        \end{subfigure}
        \begin{subfigure}[b]{0.49\textwidth}
                \centering
                \includegraphics[width=\textwidth]{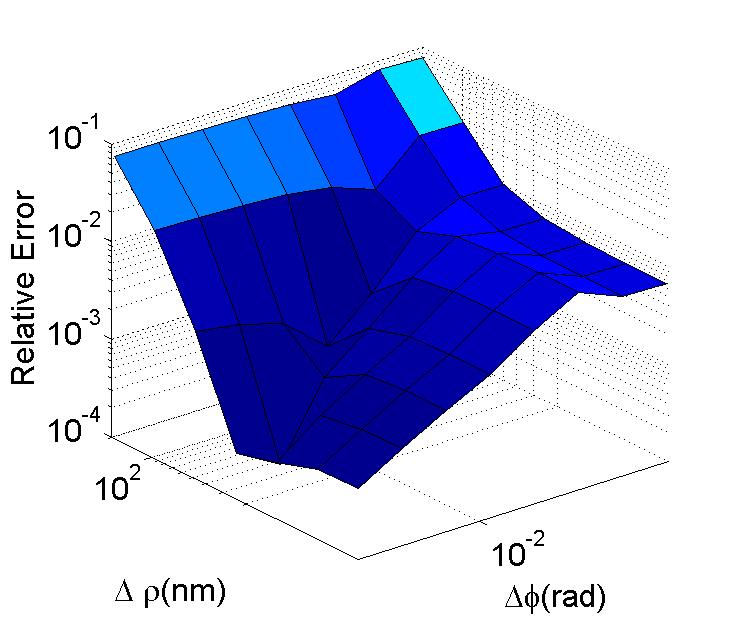}
                \caption{}
         \end{subfigure}
         \begin{subfigure}[b]{0.47\textwidth}
                \centering
                \includegraphics[width=\textwidth]{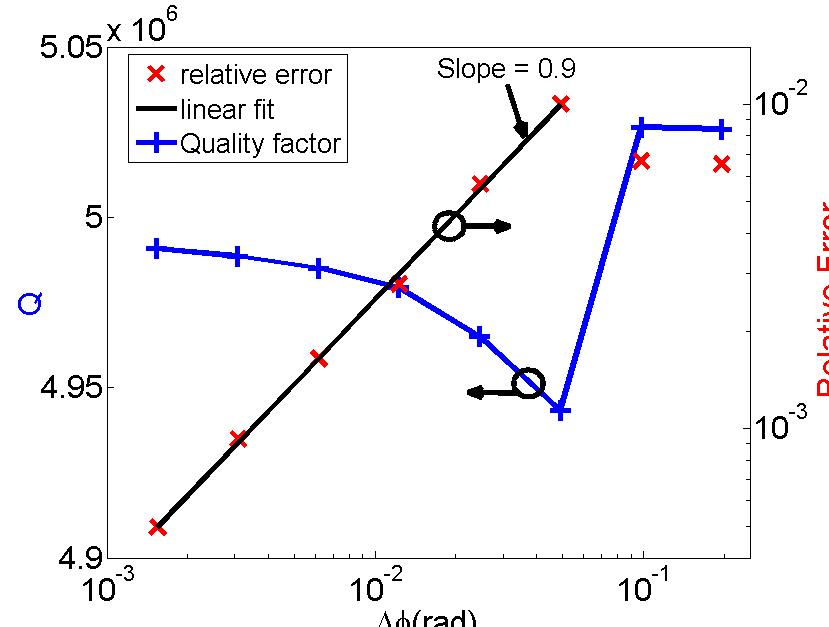}
                \caption{}
         \end{subfigure}
         \begin{subfigure}[b]{0.49\textwidth}
                \centering
                \includegraphics[width=\textwidth]{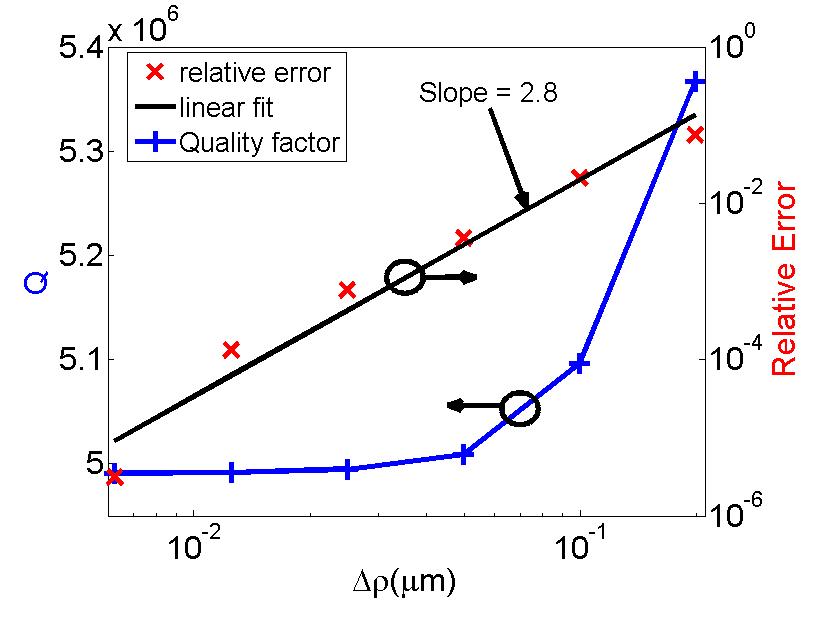}
                \caption{}
         \end{subfigure}
        \caption{(a) Quality factor vs. grid spacings, (b) its relative error vs. grid spacings, (c) variations for the cross section at $\Delta \rho = 3.1~\text{nm}$, and (d) variations for the cross section at $\Delta\phi = 0.0015~\text{radians}$. In (c), the last two points are omitted for the line of best fit.}
        \label{Q_vs_grid}
\end{figure}

We further computed the quality factor $Q$ for different grid schemes according to $Q = 2 \pi m_r P(\phi=0)/(P(\phi=0)-P(\phi=2\pi))$. $P(\phi)$ is the total power at an azimuthal angle $\phi$. Fig.~\ref{Q_vs_grid}(a) is the plot of quality factor vs. radial and azimuthal grid spacings. The computation window is set to $ 20~\mu m$ in the ${\hat \rho}\text{ direction}$ and $\pi$ in the ${\hat \phi}\text{ direction}$. As is depicted, Q converges from $5.4{\times}10^6$ towards $4.99{\times}10^6$ by reducing the grid spacing from $200~\text{nm}$ to $3.1~\text{nm}$ in the radial direction and $0.196~\text{radians}$ to $0.0015~\text{radians}$ in the azimuthal direction. We also computed the relative error by adopting the Q calculated by Richardson extrapolation as a reference value. As seen in Fig.~\ref{Q_vs_grid}(b) this was reduced to $5 \times 10^{-4}$ from $8 \times 10^{-2}$, accordingly. Furthermore, in Fig.~\ref{Q_vs_grid}(c) we plot the Q as well as the relative error as a function of azimuthal grid spacing $\Delta \phi$ by setting $\Delta\rho=3.1~\text{nm}$. In Fig.~\ref{Q_vs_grid}(d) the same quantities are plotted as a function of radial grid spacing $\Delta \rho$. The least square fits to the relative error curves indicate a convergence rate of $0.9$ in the azimuthal direction and $2.8$ in the radial direction. Clearly, this suggests that faster convergence is attainable by adopting higher-order finite-difference schemes.

\begin{figure}
   \centering
   \includegraphics[width=0.55\textwidth]{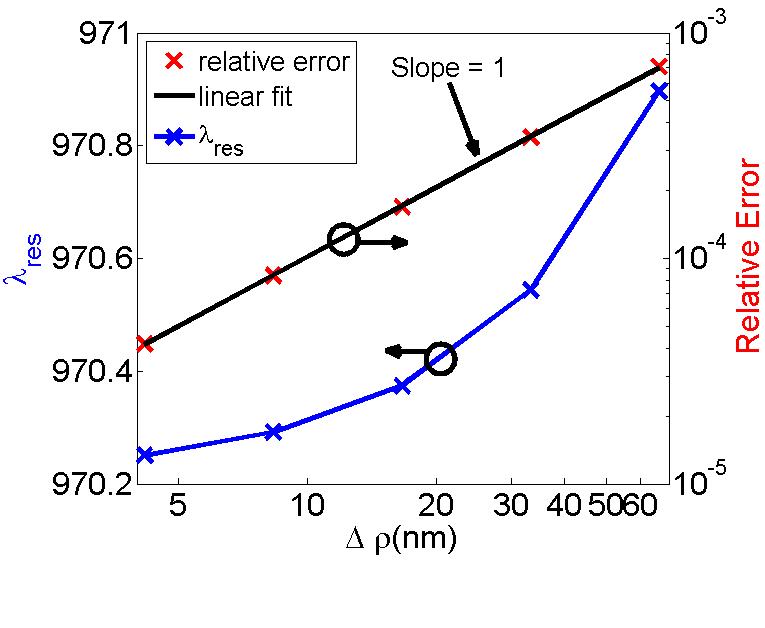}
    \caption{Resonance wavelength of the ring resonator and its relevant error vs. grid size in the ${\hat \rho}$ direction.}
    \label{l_res}
\end{figure}

In Fig.~\ref{l_res}, we plot the resonance wavelength (blue cross markers) and corresponding relative error (red cross markers) vs. radial grid spacing $\Delta \rho$ by setting $\Delta \phi=0.003~\text{radians}$. Here, the resonance wavelength $\lambda_{res}$ is obtained from the round trip total phase shift $\Delta \Phi$ of the electrical field computed from the wavelength $\lambda$ adopted in the BPM according to $\lambda_{res}=\frac{\Delta \Phi \lambda}{m_r}$, where the resonance wavelength $\lambda_{res}=970.21$ nm calculated via Richardson extrapolation is used as a reference. A least square fit to the relative error indicates a $O(\Delta \rho)$ convergence rate.

\begin{figure}
   \centering
   \includegraphics[width=1\textwidth]{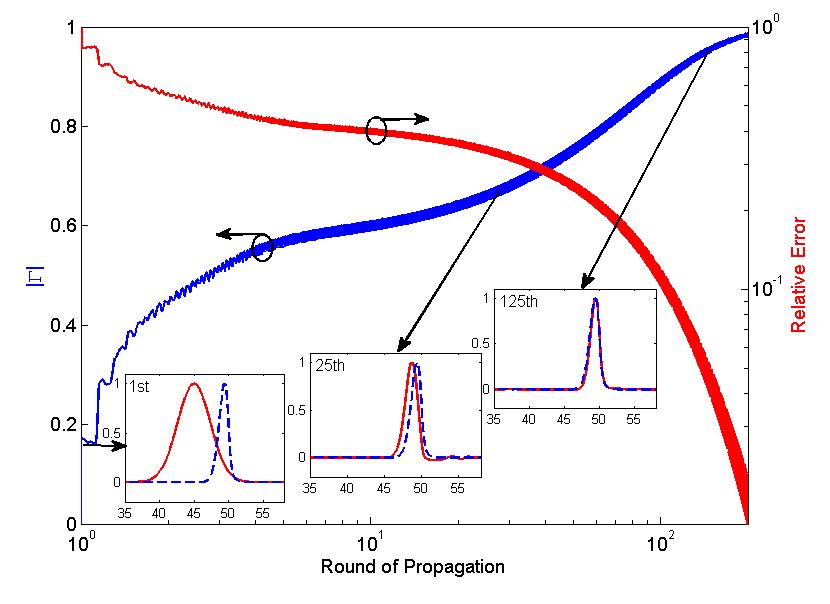}\\
   \caption{Overlap factor between the output profile and mode profile as well as its relative error vs. round trip number. The mode profile (blue) and output profile (red) at the 1$^\text{st}$, 25$^\text{th}$, and 125$^\text{th}$ rounds of propagation are plotted in insets from left to right.}\label{overlap}
\end{figure}

To further demonstrate the validity of the formulation, we launched an arbitrary Gaussian field at the input of the same structure as illustrated in the insets (red curve) of Fig.~\ref{overlap}.  For comparison, we also plotted the fundamental mode profile obtained by the mode solver and displayed it as the blue curve in the insets. As shown in the insets of Fig.~\ref{overlap}, after propagating $1$, $25$, and $125$ rounds in the resonator from left to right, the circulating field distribution gradually evolves into the mode profile. To quantitatively analyze the field evolution, we define a normalized overlap factor ${\hat \Gamma}=\langle \hat \psi_o\mid\hat\psi_{i}\rangle$. $\hat\psi_i$ is the normalized mode profile and $\hat\psi_o$ is the normalized output field profile after each round trip of propagation. The magnitude of the overlap factor and its departure from unity are respectively plotted in Fig.~\ref{overlap} as blue and red curves. As is shown, the overlap factor reaches $0.99$ at the $200^{\text{th}}$ round trip.

\begin{figure}[H]
   \centering
   \includegraphics[width=0.7\textwidth]{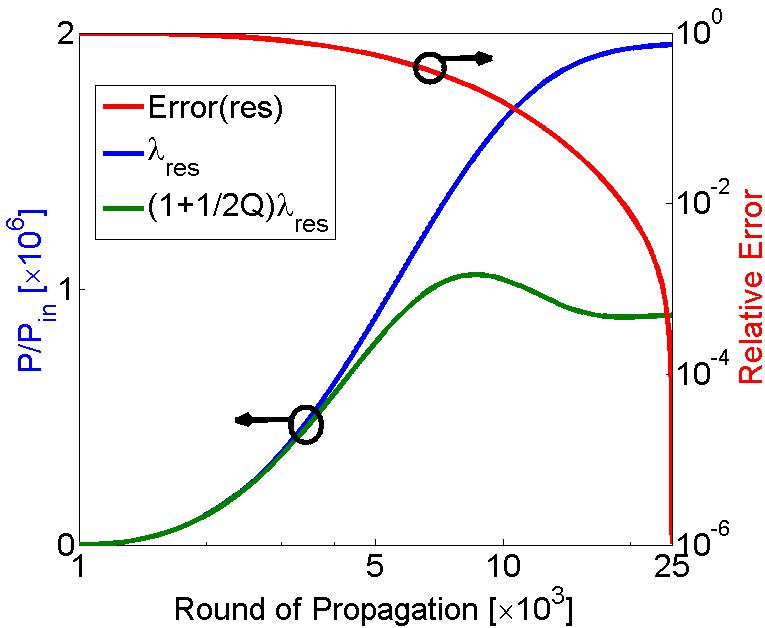}
   \caption{Power at the output of the ring after each round trip normalized to the first round power for resonance (blue), for $\lambda_{res}(1+\frac{1}{2Q})$ (green), and the resonance power's relative error (red).}
   \label{p_built_up}
\end{figure}

In Fig.~\ref{p_built_up}, we excited the cavity in continuous wave (CW) mode and plot the accumulated power normalized to the input power at the input of the cavity $P/P_{in}$ as a function of the number of rounds light circulated in the cavity. As illustrated, $P/P_{in}$ saturates to $1.95 \times 10^6$ (blue curve) when resonance occurs. The saturation power is in excellent agreement with the theoretic prediction of  $P_{sat}/P_{in}=\frac{Q^2}{\pi m_r}$ that can be obtained from a mode solver with a relative error around $10^{-6}$ after circulating more than $25,000$ rounds, as indicated in the red curve. We further plotted the total power when the input light wavelength had reached the full wave at half maximum point $(\lambda_{FWHM}=(1+1/2Q)\lambda_{res}, \text{ i.e. the green curve})$. As expected, the power saturated to half of that corresponding to resonance.

\begin{figure}[H]
        \centering
        \begin{subfigure}[b]{.49\textwidth}
                \centering
                \includegraphics[width=\textwidth]{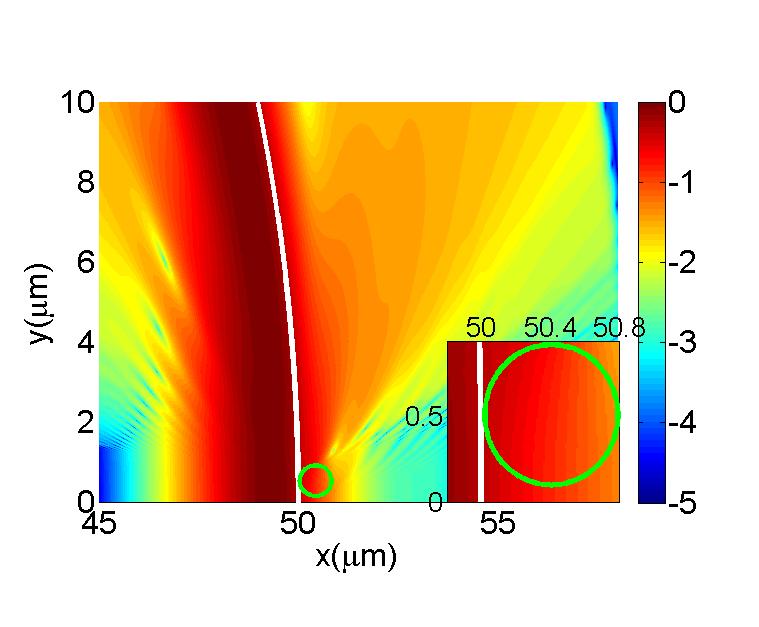}
                \caption{}
        \end{subfigure}
        \begin{subfigure}[b]{0.41\textwidth}
                \centering
                \includegraphics[width=\textwidth]{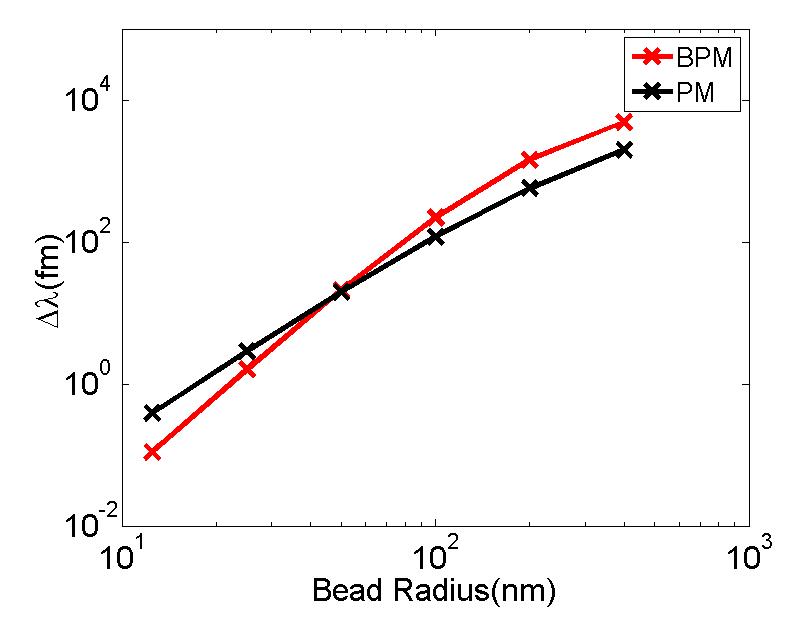}
                \caption{}
         \end{subfigure}
        \caption{(a) Field intensity (in logarithmic scale) over a slice of the ring, where the $400$-$\text{nm}$ bead is located. The inset shows the field distribution inside the bead. (b) Resonance wavelength shift vs. nanobead radius, for $\lambda = 970~\text{nm}$, calculated by the BPM method (red) and perturbation method (black).}
        \label{bead}
\end{figure}

Next, we applied the BPM to whispering-gallery mode nanodetection modelling\cite{Lu_Vahala_High_Sensitivity, biosensing_Vollmer_Arnold_Whispering_molecules}. Here we simulated the induced resonance wavelength shift caused by single polystyrene beads binding to the surface of a silica microtoroid immersed in water\cite{Lu_Vahala_High_Sensitivity}. We modeled the 3D microtoroid and bead structure in 2D using the effective index method (EIM)\cite{chiang_edd_indx}.  The field evolution across a $400$-$\text{nm}$ radius bead attached to the cavity is displayed in Fig.~\ref{bead}(a), while a zoomed-in plot of field distribution around the bead is displayed as an inset. We further obtained the resonance wavelength shift from the additional phase shift of the electrical field attributed to the bead (i.e. $\Delta \lambda_{res}=\frac{\Delta \Phi}{2 \pi m}\lambda_{res}$) and plotted them as a function of bead radius in Fig.~\ref{bead}(b) as red cross markers. For comparison, we also plotted the corresponding shift predicted by the perturbation method\cite{perturbation}. As shown, both are in good agreement aside from the fact that there is a identifiable departure due to the 2D simplification with EIM. We believe that a three-dimensional full wave beam propagation method should model the shift with higher accuracy.

\begin{figure}[H]
    \centering
    \includegraphics[width=\textwidth]{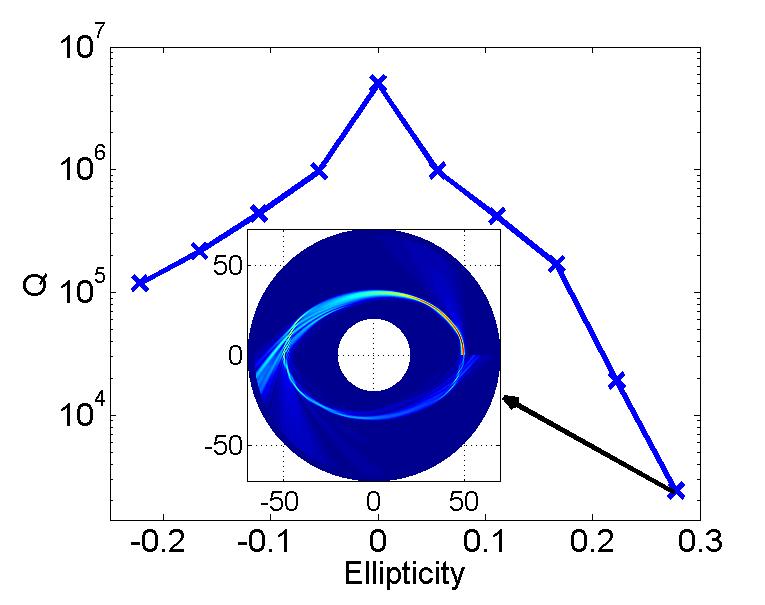}
    \caption{Quality factor vs. ellipticity. The inset shows the field intensity for $R_{0} = 45~ \mu m$ and $R_{90} = 32.5~ \mu m$.}
    \label{ellipticity}
\end{figure}

Finally, we applied the BPM to model the ellipticity effect. Setting the major radius at $\phi = 0$ to $45~\mu m$ and sweeping the other radius from $32.5~ \mu m$ to $55~ \mu m$, we calculated the quality factor as explained before using BPM. The window size is set to $50~ \mu m$ with $2201$ grids in the ${\hat \rho}~\text{ direction}$ and $2\pi~\text{radians}$ with $3000$ grids in the ${\hat \phi}\text{ direction}$. We excited the ring with its fundamental mode for a $45$-$\mu m$ major radius and $10$-$\mu m$ minor diameter ring resonator. Fig.~\ref{ellipticity} shows the quality factor as a function of ellipticity, in which the ellipticity is defined as
\begin{equation}
e = 1- \frac{R_{90}}{R_0}
\end{equation}
where $R_0 = 45~ \mu m$ is the fixed radius and $R_{90}$ is the variable radius for $\phi = 90^\circ$. The inset in Fig.~\ref{ellipticity} shows the field intensity for the extreme case of $R_{90} = 32.5~ \mu m$.

\section{Conclusion}
In conclusion, we implemented a 2D Finite-Difference Beam Propagation Method for simulating light propagation in whispering-gallery mode microcavities. With this method, we demonstrate the calculation of key optical properties such as resonance wavelength, quality factor in cases where the azimuthal symmetry is absent due to singular perturbations from nano particles, and ellipticity of the cavity. The field scattering that arises from asymmetry is clearly visible from our simulation. Such properties are not readily obtainable through first-order perturbations on whispering-gallery modes as traditionally treated by other groups. Therefore, cylindrical BPM addresses the need for in-depth study of areas such as biosensing with WGM's where azimuthal symmetry is perturbed. A full vectorial three-dimensional BPM approach will be investigated in future research.


\begin{thebibliography}{10}
\newcommand{\enquote}[1]{``#1''}

\bibitem{toroid_Gorodetsky_Ilchenko_Ultimate_Microsphere}
M.~L. Gorodetsky, A.~A. Savchenkov, and V.~S. Ilchenko, \enquote{Ultimate {Q}
  of optical microsphere resonators,} Optics Letters \textbf{21}, 453--455
  (1996).

\bibitem{toroid_Armani_Ultra}
D.~K. Armani, T.~J. Kippenberg, S.~M. Spillane, and K.~J. Vahala,
  \enquote{Ultra-high-{Q} toroid microcavity on a chip,} Nature \textbf{421},
  925--928 (2003).

\bibitem{Vollmer_Arnold_Protein_Microcavity}
F.~Vollmer, D.~Braun, A.~Libchaber, M.~Khoshsima, I.~Teraoka, and S.~Arnold,
  \enquote{Protein detection by optical shift of a resonant microcavity,}
  Applied Physics Letters \textbf{80}, 4057--4059 (2002).

\bibitem{biosensing_Vollmer_Arnold_Whispering_molecules}
F.~Vollmer and S.~Arnold, \enquote{Whispering-gallery-mode biosensing:
  labelfree detection down to single molecules,} Nat. Methods \textbf{5},
  591--596 (2008).

\bibitem{Lu_Vahala_High_Sensitivity}
T.~Lu, H.~Lee, T.~Chen, S.~Herchak, J.-H. Kim, S.~E. Fraser, and K.~Vahala,
  \enquote{High sensitivity nanoparticle detection using optical
  microcavities,} PNAS \textbf{108}, 5976--5979 (2011).

\bibitem{Dominguez_Martorell_Whispering_molecules}
J.~Dominguez-Juarez, G.~Kozyreff, and J.~Martorell, \enquote{Whispering gallery
  microresonators for second harmonic light generation from a low number of
  small molecules,} Nautre Communications \textbf{2}, 1--8 (2010).

\bibitem{Knittel_Bowen_Interferometric_biosensors}
J.~Knittel, T.~G. McRae, K.~H. Lee, and W.~P. Bowen, \enquote{Interferometric
  detection of mode splitting for whispering gallery mode biosensors,} Applied
  Physics Letters \textbf{97}, 1--3 (2010).

\bibitem{Sun_Fan_Optical_Sensing}
Y.~Sun and X.~Fan, \enquote{Optical ring resonators for biochemical and
  chemical sensing,} Anal bioanal Chem \textbf{399}, 205--211 (2011).

\bibitem{Spillane_Vahala_Ultralow_Microcavity}
S.~Spillane, T.~J. Kippenberg, and K.~J. Vahala, \enquote{Ultralow-threshold
  raman laser using a spherical dielectric mcirocavity,} Nature \textbf{415},
  621--623 (2002).

\bibitem{Min_Vahala_Compact_Laser}
B.~Min, T.~J. Kippenberg, and K.~J. Vahala, \enquote{Compact, fiber-compatible,
  cascaded raman laser,} Optics Letters \textbf{28}, 1507--1509 (2003).

\bibitem{Cai_Vahala_Highly_Laser}
M.~Cai and K.~J. Vahala, \enquote{Highly efficient hybrid fiber taper coupled
  microsphere laser,} Optics Letters \textbf{26}, 884--886 (2001).

\bibitem{toroid_Polman_Ultralow}
A.~Polman, B.~Min, J.~Kalkman, T.~J. Kippenberg, and K.~J. Vahala,
  \enquote{Ultralow-threshold erbium-implanted toroidal microlaser on silicon,}
  Applied Physics Letters \textbf{84}, 1037--1039 (2004).

\bibitem{Lu_Vahala_On_Chip_Microlaser}
T.~Lu, L.~Yang, R.~V.~A. van Loon, A.~Polman, and K.~J. Vahala,
  \enquote{On-chip green silica upconversion microlaser,} Optics Letters
  \textbf{34}, 482--484 (2009).

\bibitem{Lu_Min_Narrow_Laser}
T.~Lu, L.~Yang, T.~Carmon, and B.~Min, \enquote{A narrow-linewidth on-chip
  toroid raman laser,} IEEE J. Quantum Electron. \textbf{47}, 320--326 (2011).

\bibitem{toroid_Spillane_Ultrahigh}
S.~M. Spillane, T.~J. Kippenberg, K.~J. Vahala, K.~W. Goh, E.~Wilcut, and H.~J.
  Kimble, \enquote{Ultrahigh-{Q} toroidal microresonators for cavity quantum
  electrodynamics,} Physical Review A \textbf{71}, 013817 (2005).

\bibitem{toroid_Oxborrow_Traceable}
M.~Oxborrow, \enquote{Traceable 2-{D} finite-element simulation of the
  whispering-gallery modes of axisymmetric electromagnetic resonators,} {IEEE}
  Transactions on Microwave Theory and Techniques \textbf{55}, 1209--1218
  (2007).

\bibitem{Poon_Yariv_Matrix_Waveguides}
J.~K.~S. Poon, J.~Scheuer, S.~Mookherjea, G.~T. Paloczi, Y.~Huang, and
  A.~Yariv, \enquote{Matrix analysis of microring coupled-resonator optical
  waveguides,} Opt. Express \textbf{12}, 90--103 (2004).

\bibitem{TMM}
J.~Hong, W.~P. Huang, and T.~Makino, \enquote{On the transfer matrix method for
  distributed-feedback waveguide devices,} J. Lightwave Technol. \textbf{10},
  1860--1868 (1992).

\bibitem{Du:13}
X.~Du, S.~Vincent, and T.~Lu, \enquote{Full-vectorial whispering-gallery-mode
  cavity analysis,} Opt. Express \textbf{21}, 22012--22022 (2013).

\bibitem{Feit_Fleck_Light_Fibers}
M.~D. Feit and J.~J.~A.~Fleck, \enquote{Light propagation in graded-index
  optical fibers,} Appl. Opt. \textbf{17}, 3990--3998 (1978).

\bibitem{yevick_1990}
D.~Yevick and B.~Hermansson, \enquote{Efficient beam propagation techniques,}
  IEEE J. Quantum Electron. \textbf{26}, 109 (1990).

\bibitem{yevick1983}
J.~Saijonmaa and D.~Yevick, \enquote{Beam-propagation analysis of loss in bent
  optical waveguides and fibers,} J. Opt. Soc. Am. \textbf{73}, 1785 (1983).

\bibitem{huang-march92}
W.~Huang, C.~Xu, S.-T. Chu, and S.~K. Chaudhuri, \enquote{The finite-difference
  vector beam propagation method: Analysis and assessment,} J. Lightwave
  Technol. \textbf{10}, 295--305 (1992).

\bibitem{vanroey1981}
J.~V. Roey, J.~van~der Donk, and P.~E. Lagasse, \enquote{Beam-propagation
  method: analysis and assessment,} J. Opt. Soc. Am. \textbf{71}, 803--810
  (1981).

\bibitem{huang1992}
W.~Huang, C.~Xu, and S.~Chaudhuri, \enquote{A finite-difference vector beam
  propagation method for three-dimensional waveguide structures,} IEEE Photon.
  Technol. Lett. \textbf{4}, 148--151 (1992).

\bibitem{yevick1984}
B.~Hermansson and D.~Yevick, \enquote{Propagating-beam-method analysis of
  two-dimensional microlenses and three-dimensional taper structures,} Opt.
  Soc. Am. A \textbf{1}, 663--671 (1984).

\bibitem{Osgood}
H.~Rao, R.~Scarmozzino, and J.~Richard M.~Osgood, \enquote{A bidirectional beam
  propagation method for multiple dielectric interfaces,} IEEE Photon. Technol.
  Lett. \textbf{11}, 830--832 (1999).

\bibitem{Gopinath}
R.~Scarmozzino, A.~Gopinath, R.~Pregla, and S.~Helfert, \enquote{Numerical
  techniques for modeling guided-wave photonic devices,} IEEE J. Sel. Top.
  Quantum Electron. \textbf{6}, 150--162 (2000).

\bibitem{FE-BPM_bangiladish}
A.~Ahmed, R.~Koya, O.~Wada, M.~Wang, and R.~Koga, \enquote{Eigenmode analysis
  of whispering gallery mmode of pillbox-type optical resonators utilizing the
  fe-bpm formulation,} IEICE Tans. Electron. \textbf{E78-C}, 1638 (1995).

\bibitem{Yang_Boundary_new}
W.~Yang and A.~Gopinath, \enquote{A boundary integral method for propagation
  problems in integrated optical structures,} IEEE Photon. Technol. Lett.
  \textbf{7}, 777--779 (1995).

\bibitem{pmd_Lu_Boundary}
T.~Lu and D.~Yevick, \enquote{Boundary element analysis of dielectric
  waveguides,} Journal of Optical Society of America A \textbf{19}, 1197 --
  1206 (2002).

\bibitem{pmd_Lu_Comparative}
T.~Lu and D.~Yevick, \enquote{Comparative evaluation of a novel series
  approximation for electromagnetic fields at dielectric corners with boundary
  element method applications,} {IEEE/OSA} Journal of Lightwave Technology
  \textbf{22}, 1426--1432 (2004).

\bibitem{pmd_Lu_Vectorial}
T.~Lu and D.~Yevick, \enquote{A vectorial boundary element method analysis of
  integrated optical waveguides,} {IEEE/OSA} Journal of Lightwave Technology
  \textbf{21}, 1793--1807 (2003).

\bibitem{pmd_Deng_Nonunitarity_new}
H.~Deng and D.~Yevick, \enquote{The nonunitarity of finite-element beam
  propagation algorithms,} IEEE Photon. Technol. Lett. \textbf{17}, 1429--1431
  (2005).

\bibitem{FDTD_waterloo}
S.-T. Chu and S.~Chaudhuri, \enquote{A finite-difference time-domain method for
  the design and analysis of guided-wave optical structures,} J. Lightwave
  Technol. \textbf{7}, 2033--2038 (1989).

\bibitem{mode1996}
M.~Reed, T.~M. Benson, P.~C. Kendall, and P.~Sewell,
  \enquote{Antireflectioncoated angled facet design,} Proc. Inst. Elect. Eng.
  \textbf{143}, 214--220 (1996).

\bibitem{PML_Berenger1994}
Berenger and Jean-Pierre, \enquote{A perfectly matched layer for the absorption
  of electromagnetic waves,} J. Comput. Phys. \textbf{114}, 185--200 (1994).

\bibitem{PML_Huang}
W.~P. Huang, C.~L. Xu, W.~Lui, and K.~Yokoyama, \enquote{The perfectly matched
  layer ({PML}) boundary condition for the beam propagation method,} IEEE
  Photon. Technol. Lett. \textbf{8}, 649--651 (1996).

\bibitem{hadley2002}
G.~R. Hadley, \enquote{High-accuracy finite-difference equations for dielectric
  waveguide analysis {I}: Uniform regions and dielectric interfaces,} J.
  Lightwave Technol. \textbf{20}, 1210--1218 (2002).

\bibitem{Wide_angle_Hadley-Pade}
G.~R. Hadley, \enquote{Wide-angle beam propagation using pade approximant
  operators,} Opt. Lett. \textbf{17}, 1426--1428 (1992).

\bibitem{cyl-FD-BPM}
S.~Lidgate, P.~Sewell, and T.~Benson, \enquote{Conformal mapping: limitations
  for waveguide bend analysis,} Sci. Meas. Technol. \textbf{149}, 262--266
  (2002).

\bibitem{rivera1995}
M.~Rivera, \enquote{A finite difference {BPM} analysis of bent dielectric
  waveguides,} J. Lightwave Technol. \textbf{13}, 233 (1995).

\bibitem{3D-curved}
H.~Deng, G.~H. Jin, J.~Harari, J.~P. Vilcot, and D.~Decoster,
  \enquote{Investigation of 3-{D} semivectorial finite-difference beam
  propagation method for bent waveguides,} J. Lightwave Technol. \textbf{16},
  915--922 (1998).

\bibitem{krause2011}
M.~Krause, \enquote{Finite-difference mode solver for curved waveguides with
  angled and curved dielectric interfaces,} J. Lightwave Technol. \textbf{29},
  691--699 (2011).

\bibitem{silica_refract}
I.~H. Malitson, \enquote{Interspecimen comparison of the refractive index of
  fused silica,} J. Opt. Soc. Am. \textbf{55}, 1205--1208 (1965).

\bibitem{refractiv_imaginary}
R.~Kitamura, L.~Pilon, and M.~Jonasz, \enquote{Optical constants of silica
  glass from extreme ultraviolet to far infrared at near room temperature,}
  Appl. Opt. \textbf{46}, 8118--8133 (2007).

\bibitem{Water_refract}
G.~M. Hale and M.~R. Querry, \enquote{Optical constants of water in the 200-nm
  to 200-Âµm wavelength region,} Appl. Opt. \textbf{12}, 555--563 (1973).

\bibitem{chiang_edd_indx}
K.~S. Chiang, \enquote{Performance of the effective-index method for the
  analysis of dielectric waveguides,} Opt. Lett. \textbf{16}, 714--716 (1991).

\bibitem{perturbation}
R.~A. Waldron, M.A., and A.Inst.P., \enquote{Perturbation theory of resonant
  cavities,} Proceedings of the IEE - Part C: Monographs \textbf{107}, 272--274
  (1960).

\end{thebibliography}
\end{document}